\documentclass[pra,twocolumn,showpacs,amsmath,amssymb]{revtex4-1}

\usepackage{psfrag,graphicx}
\usepackage{amsfonts,amssymb,amsmath}      

\usepackage{graphicx}
\usepackage{dcolumn}
\usepackage{bm}
\usepackage{hyperref}
\usepackage{color}
\usepackage{epstopdf}

\newcommand{\ket}[1]{|{#1}\rangle}
\newcommand{\bra}[1]{\langle{#1}|}
\newcommand{\nn}{\nonumber\\}

\begin{document}

\preprint{APS/123-QED}

\title{Loss-resistant unambiguous phase measurement}

\author{Hossein T. Dinani}
\author{Dominic W. Berry}
\affiliation{Department of Physics and Astronomy, Macquarie University, Sydney, NSW 2109, Australia}


\date{\today}

\begin{abstract}
Entangled multi-photon states have the potential to provide improved measurement accuracy, but are sensitive to photon loss.
It is possible to calculate ideal loss-resistant states that maximize the Fisher information, but it is unclear how these could be experimentally generated.
Here we propose a set of states that can be obtained by processing the output from parametric down-conversion.
Although these states are not optimal, they provide performance very close to that of optimal states for a range of parameters.
Moreover, we show how to use sequences of such states in order to obtain an unambiguous phase measurement that beats the standard quantum limit.
We consider the optimization of parameters in order to minimize the final phase variance, and find that the optimum parameters are different from those that maximize the Fisher information.
\end{abstract}

\maketitle

\section{Introduction}
Precision measurement is vital to many areas of science and technology, and the most accurate measurement techniques are typically based on interferometry.
The most accurate distance measurements are used in gravitational wave detectors, where advances in accuracy provide an entirely new way of observing the universe \cite{LIGO}. 
Quantum mechanics imposes limits on the possible accuracy, but by taking advantage of the exotic features of quantum mechanics it is possible to obtain far better accuracy than would otherwise be possible \cite{Caves,Holland,Giovannetti,Giovannetti11}.

In particular, in optical interferometers one could use an effectively classical approach, where the photons are independent of each other.
Using $N$ independent photons, the phase uncertainty scales as $1/\sqrt N$ -- a scaling known as the standard quantum limit (SQL).
The SQL can be beaten using nonclassical states where the photons are correlated; for example by using squeezed states \cite{Caves} or two-mode entangled states \cite{Kok}.
This leads to an ultimate limit of $1/N$ for the phase uncertainty, often called the Heisenberg limit \cite{Holland,Ou,Braun,Zwierz,Gio,Hall12}.

One particular type of nonclassical state that has been widely considered is the NOON state $\left({\left| {N,0} \right\rangle  + \left| {0,N} \right\rangle}\right)/\sqrt 2$ \cite{Barry,LeeKokJD,Dowling}.
These states give Heisenberg scaling for the phase uncertainty, in the sense that they give phase information on a scale of $1/N$.
However, the phase information obtained is highly ambiguous, and additional phase information is needed to resolve this ambiguity.
This can be achieved, for example, by using sequences of NOON states with different values of $N$ \cite{HigginsN,DominicPRA9,NJPHiggins,Pezze}.

NOON states are also highly sensitive to photon loss \cite{Rubin}; loss of even one photon destroys all phase information.
On the other hand, combining single photon states with low-photon-number NOON states can beat the SQL in the presence of loss \cite{Cooper}.
More generally, one can consider states that are optimized to be robust to photon loss \cite{Dorner,Dobrzanski,Knott,Kacprowicz}.
Nevertheless, even using optimized states, when there is photon loss the lower limit to the phase uncertainty is just a constant factor improvement over the SQL in the limit of large $N$  \cite{Kolo10,Knysh11,Escher,Dobrzanskinc}.

In considering optimized states for two-mode interferometry, they can be taken to be of the form $\sum_{k = 0}^N \psi_k\ket{N - k,k}$.
The total photon number can be taken to be a single value $N$; there is no advantage to using a superposition over different total photon numbers, because the detection process will destroy any such superposition.
In the case of NOON states, one has only two nonzero values of $\psi_k$, whereas general optimized states would, in general, require all values of $\psi_k$ to be nonzero.
This raises a challenging state engineering problem \cite{Lee02,Fiur,Pryde,vanMeter}.
Such a general state could, for example, be produced by using $N$ independent single photons \cite{Zou}.

However, the standard method to produce single photons is to use down-conversion, and post-select on detection of a single photon to obtain a single photon in the other output.
This means that such a scheme would be very wasteful, because $2N$ photons would need to be produced to obtain a state with $N$ photons.
Here we consider the question of the types of states that can be produced by using all photons output by the down-conversion.
Our two-photon states are similar to the states proposed in \cite{Kacprowicz}, but the scheme proposed here is also able to produce states with larger numbers of photons.
Moreover, we consider how to use sequences of such states with different total numbers of photons in order to obtain unambiguous phase estimates, in much the same way as has been done for NOON states \cite{HigginsN,DominicPRA9,Pezze}.

\section{The state preparation scheme}\label{section-statepreparation}
\begin{figure}[t]
\centering
\includegraphics[scale=0.82]{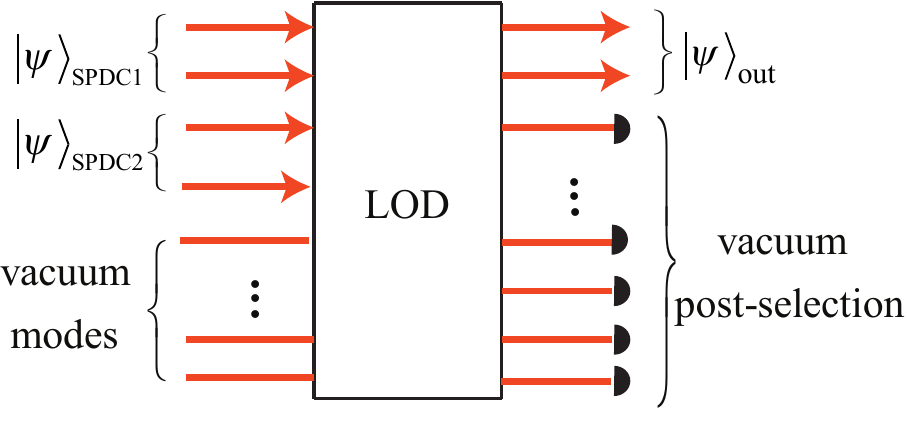}
\caption{A linear optical device (LOD) with two SPDC states and vacuum modes as input.
$\ket\psi_{\rm out}$ is the desired output state, and there is post-selection on vacuum in the other modes.
The two SPDC input states can be simplified to one.}
\label{LOD}
\end{figure}

We start with an $M$-port linear optical device (LOD) shown in Fig.~\ref{LOD}.
An LOD can be decomposed into a triangular array of beam splitters and phase shifters \cite{Reck}.
Any $M$ port LOD can be represented  by an $M \times M$ unitary matrix $U$ with elements $U_{ij}$.
It transforms input photon creation operators $\hat a^{\dagger}_i$ to the output creation operators $\hat b^{\dagger}_i$ as
\begin{equation}\label{lodtrans}
\hat a_i^\dag  = \sum_{j=1}^M U_{ij}\hat b_{j}^\dag.
\end{equation}

Our aim here is to propose an LOD scheme to generate two-mode multi-photon entangled states in the output which are resistant to photon loss.
The most general two mode $N$-photon state can be written as
\begin{equation}
\left| \psi  \right\rangle _{\rm out} = \sum\limits_{k = 0}^N \psi_k\frac{1}{\sqrt {k!(N-k)!}}\left( \hat b_{1}^\dag\right)^{N - k}\left( \hat b_{2}^\dag \right)^k\ket{0,0} 
\end{equation}
where the quantities $\psi_k$ are the probability amplitudes of the photon number components.
Optimal states for phase estimation are symmetric between the paths \cite{Hoffmann}, so we aim to have $\psi_k = \psi_{N-k}$ in our output state. 
We post-select vacuum in the other $M-2$ output modes of the LOD in order to maximize the number of photons in the output.

Spontaneous parametric down-conversion (SPDC) is the most common source of photon pairs used in experiments.
In order to make our state preparation scheme experimentally feasible, we consider the output from SPDC as the input to the LOD.
The SPDC state can be written as \cite{Ou07}
\begin{equation}
{\left| \psi  \right\rangle _\text{SPDC}} \propto  {\left| 0 \right\rangle _s}{\left| 0 \right\rangle _i} + \xi {\left| 1 \right\rangle _s}{\left| 1 \right\rangle _i} + {\xi ^2}{\left| 2 \right\rangle _s}{\left| 2 \right\rangle _i} + {\xi ^3}{\left| 3 \right\rangle _s}{\left| 3 \right\rangle _i} + ...
\end{equation}
where the subscripts $s$ and $i$ stand for signal and idler; these will be omitted from this point on for brevity.
We also use the proportional to symbol to indicate that a normalization constant has been omitted.
The quantity $\xi$ depends on the interaction time between the optical nonlinear crystal and pump laser, the strength of the nonlinearity in the crystal and the power of the pump laser.
Probability amplitudes of each term in the above SPDC state can be controlled by the power of the pump laser.

If we input the LOD with one SPDC source, recording a total of $N=2n$ photons in the output post-selects only the dual Fock state $\left| n,n\right\rangle$ as input
\begin{equation}\label{dualF}
\left| {n,n} \right\rangle  = \frac{1}{{n!}}{\left( {\hat a_1^\dag } \right)^n}{\left( {\hat a_2^\dag } \right)^n}\left| {0,0} \right\rangle .
\end{equation}
If vacuum is post-selected in $M-2$ of the output modes (so the $2n$ photons are detected in the first two modes),  the output state can be represented by removing all terms containing $\hat b^{\dagger}_i$ for $i \ge 3$.
Transforming creation operators in Eq.~\eqref{dualF} by Eq.~\eqref{lodtrans}, and omitting $\hat b^{\dagger}_i$ with $i \ge 3$ gives the output state in the form
\begin{equation}\label{twomodeg}
\ket\psi_{\rm out} = \frac 1{n!} \left[ \chi_1\left(\hat b_1^\dag\right)^2 + \chi_2\hat b_1^\dag \hat b_2^\dag  + \chi_3\left(\hat b_2^\dag \right)^2 \right]^n\ket{0,0}
\end{equation}
where $\chi_1,\, \chi_2, \, \chi_3$ are the factors determined by the unitary matrix $U$.

Alternatively, consider the case where there were two SPDC sources used, as in Fig.~\ref{lodtrans}.
Allowing different factors $\xi_1$ and $\xi_2$, the input state can be written as
\begin{align}\label{twoSPDC}
\ket\psi_{\rm in} &\propto \left( \ket{0,0} + \xi_1 \ket{1,1} + \xi_1^2\ket{2,2} + \xi_1^3\ket{3,3}+... \right)\nn
& \quad \otimes \left( \ket{0,0} + \xi_2 \ket{1,1} + \xi_2^2\ket{2,2} + \xi_2^3\ket{3,3}+... \right) \nn
&\propto {e^{\xi_1 \hat a_1^\dag \hat a_2^\dag  + \xi_2 \hat a_3^\dag \hat a_4^\dag }}\ket{0,0}\ket{0,0} \nn
&= \sum_{n=0}^\infty \frac 1{n!}\left( \xi_1 \hat a_1^\dag \hat a_2^\dag  + \xi_2 \hat a_3^\dag \hat a_4^\dag \right)^n \ket{0,0} \ket{0,0}.
\end{align}
Recording a total of $2n$ photons in the output modes of LOD post-selects the state 
\begin{equation}
\ket\psi_{\rm in} \propto \frac 1{n!} \left( \xi_1 \hat a_1^\dag \hat a_2^\dag  + \xi_2 \hat a_3^\dag \hat a_4^\dag \right)^n\ket{0,0} \ket{0,0}
\end{equation}
as the input state.
If vacuum is post-selected in all but two of the output modes, we again obtain the state given in Eq.~\eqref{twomodeg}.
Similarly, if we were to consider an arbitrary number of SPDC sources, we would again obtain the same form of state.
Therefore there is no advantage to considering larger numbers of SPDC sources, and we consider an LOD fed with only one SPDC source and vacuum to the remaining modes.
As we aim for a symmetric state we set $\chi_1=\chi_3$.
It is also convenient to denote $\chi=\chi_2/\chi_1$, which makes it clear that the states are parametrized by just one real number.

\begin{figure}[tbh]
\centering
\includegraphics[scale=0.55]{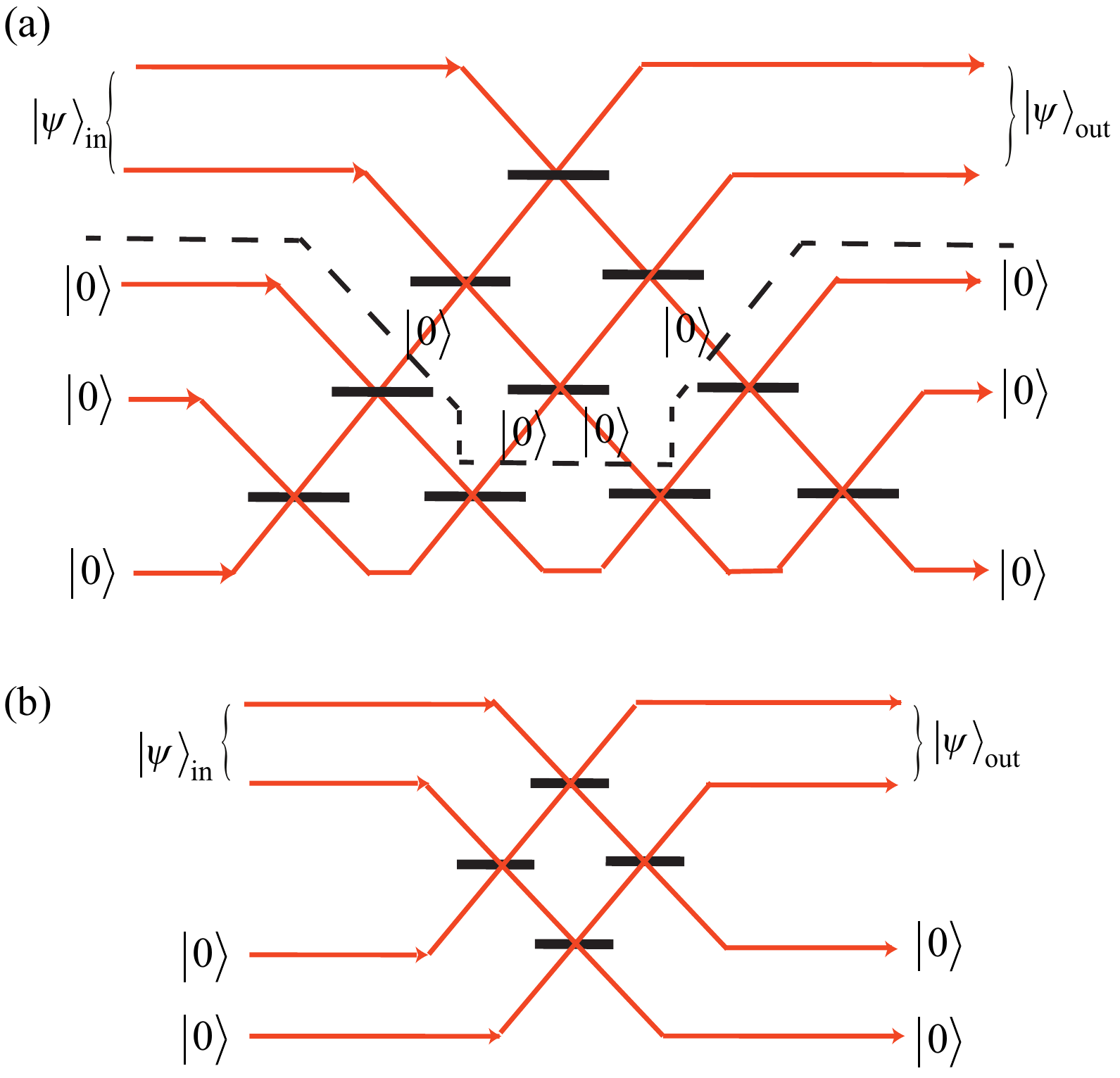}
\caption{ (a) A triangular array of beam splitters and phase shifts fed with a two mode input state $\ket\psi_{\rm in}$ and three vacuum modes, $\ket 0$.
The phase shifts are included with the beam splitters (shown as thick black lines) for simplicity.
Output photons are detected in two of the modes, $\ket\psi_{\rm out}$, and vacuum is post-selected in three of the output modes.
All the arms below the dashed line are vacuum.
(b) The five-port LOD can be simplified to a four-port one by keeping only the beam splitters above the dotted line.}
\label{53port}
\end{figure}

Thus we consider an $M$-port LOD in which all but two of the input modes are vacuum and also all but two of the output modes are post-selected as vacuum, but we are allowing a potentially large number of modes.
The scheme can be simplified to a four-port LOD in the following way.
Figure~\ref{53port}(a) shows a five-port LOD with three vacuum input modes and post-selection of vacuum at three modes in the output, with the LOD simplified to a triangular array of beam splitters and phase shifters \cite{Reck}.
It can be seen that there are six beam splitters that have vacuum input and output [those below the dashed line in Fig.~\ref{53port}(a)].
These beam splitters leave the field unchanged, and can be omitted.
Therefore the scheme can be simplified to the four-port LOD shown in Fig.~\ref{53port}(b).
Similarly, if we started with an arbitrary number of modes ($>4$), the scheme could be simplified to four modes.

\begin{figure}[t]
\centering
\includegraphics[scale=0.8]{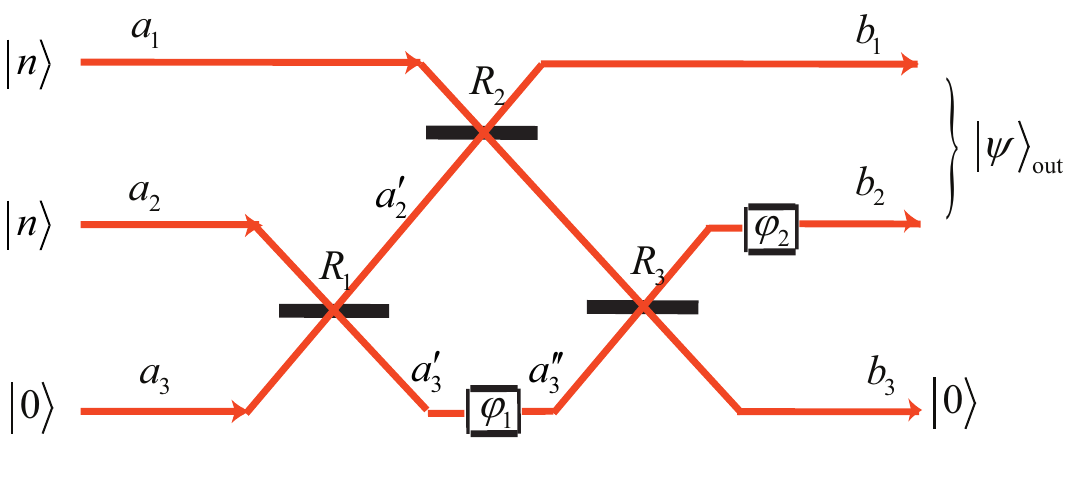}
\caption{The state preparation scheme for loss-resistant states. It is a three-port interferometer consisting of three beam splitters with reflectivities $R_1,\, R_2,\, R_3$ and two phase shifters $\varphi_1,\, \varphi_2$. The initial dual Fock state $\left| n \right\rangle \left| n \right\rangle $ gives the $2n$-photon loss-resistant state in the output, as indicated in Eq.~\eqref{lr}.}
\label{thescheme}
\end{figure}
This shows that we can simplify to four modes, but we have found an even simpler scheme using three modes, as shown in Fig.~\ref{thescheme}.
In this figure, $R_1, R_2, R_3$ are reflectivities of beam splitters and $\varphi_1, \varphi_2$ are phases of phase shifters.
We use $\hat a_1^\dag$, $\hat a_2^\dag$, $\hat a_3^\dag$ for the creation operators of the input modes and  $\hat b_1^\dag$, $\hat b_2^\dag$, $\hat b_3^\dag$ for the output modes. Labeling the modes after the first beam splitter as $a'_2$ and $a'_3$ we have
\begin{align}\label{bs}
\hat a_2^\dag  &= i\sqrt {R_1} \hat a'^\dag_2 + \sqrt {T_1} \hat a'^\dag_3 , \nn
\hat a_3^\dag  &= i\sqrt {R_1} \hat a'^\dag_3 + \sqrt {T_1} \hat a'^\dag_2
\end{align}
where $T_j=1-R_j$ is the transmissivity of the beam splitters.
The phase shifters only change the phase of the modes.
Labeling the mode after the first phase shifter as $a''_3$, the action of the first phase shifter is
\begin{equation}\label{ps}
\hat a'^\dag_3 = e^{i\varphi_1 }\hat a''^\dag_3.
\end{equation}

Inputing the state
\begin{equation}
\left| {{\psi }} \right\rangle_{\rm{in}}  = \left| {n,n,0} \right\rangle  = \frac{1}{{n!}}{\left( {\hat a_1^\dag } \right)^n}{\left( {\hat a_2^\dag } \right)^n}\left| {0,0,0} \right\rangle
\end{equation}
to the scheme, acting the beam splitters and phase shifters according to Eqs.~\eqref{bs} and \eqref{ps} and post-selecting the vacuum in the $b_3$ mode in the output (so $\hat b_3^\dag$ and its powers are replaced with zero) gives us the state
\begin{align}\label{psi4}
&\ket\psi_{\rm out} \propto \frac 1{n!} \left( i\sqrt {R_2} \hat b_1^\dag + i e^{i\varphi _2}\sqrt {T_2 R_3} \hat b_2^\dag \right)^n \left[ i\sqrt {R_1 T_2} \hat b_1^\dag \right. \nn
& \quad + \left. \hat b_2^\dag e^{i\varphi _2}\left( e^{i \varphi _1}\sqrt{T_1 T_3}  - i\sqrt{R_1 R_2 R_3} \right)\right]^n\ket{0,0}.
\end{align}

To obtain the state 
\begin{equation}\label{lr}
\ket\psi_{\rm out}  \propto \left[ \left( \hat b_1^\dag \right)^2 + \chi\hat b_1^\dag \hat b_2^\dag+\left( \hat b_2^\dag \right)^2 \right]^n\ket{0,0}
\end{equation}
with $\chi\in[0,2]$, we can use the reflectivities and phase shifts as 
\begin{align}
{\varphi _1} &= \arcsin \left( {\frac{1}{2}(\chi-1)\sqrt {2+\chi} } \right),\,\,\,{\varphi _2} = \arccos \left( {\frac{\chi}{2}} \right), \nonumber \\
{R_1} &= \frac{1}{1 + \chi},\quad {R_2} = \frac{1}{2 + \chi},\quad R_3= \frac{1}{1 + \chi}.
\end{align}
For the two-photon state, $n=1$,  the output state, Eq.~\eqref{lr}, can be expanded as 
\begin{equation}\label{N2}
{\left| \psi  \right\rangle _{{\rm{out}}}} \propto \sqrt 2 \left| {0,2} \right\rangle  + \chi \left| {1,1} \right\rangle  + \sqrt 2 \left| {2,0} \right\rangle .
\end{equation}
For the $n=2$, four-photon state, we get the following expansion 
\begin{equation}\label{N4}
\ket\psi_{\rm out} \propto \ket{0,4} + \chi \ket{1,3} + \frac{2 + \chi^2}{\sqrt 6}\ket{2,2} + \chi \ket{3,1} + \ket{4,0}.
\end{equation}
Unlike NOON states, the above states have terms other than $\left| {N,0} \right\rangle$ and $\left| {0,N} \right\rangle$.
This makes these states more resilient to photon loss; in other words, loss of a single photon does not destroy the phase sensitivity of the states.
Our proposed states still have some ambiguity in phase estimation, but by combining them with single photon states the ambiguity can be removed.
By adjusting the reflectivities of the beam splitters and phases of the phase shifters we can choose the value of $\chi$ and optimize for phase measurement with loss.

\begin{figure}[tbh]
\centering
\includegraphics[scale=0.7]{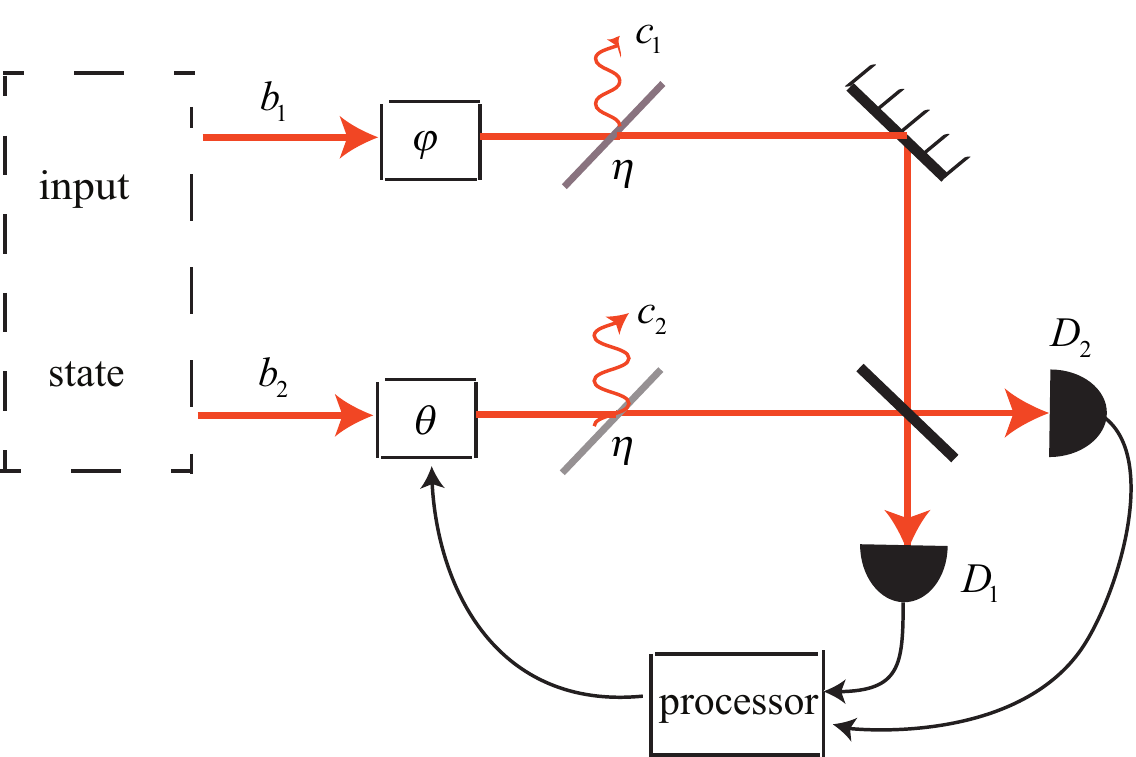}
\caption{Adaptive phase measurement scheme. $\varphi$ is the unknown phase, $\theta$ is the controlled phase and $D_1$ and $D_2$ are the photon detectors in the two outputs. The diagonal lines in the two arms are the fictitious beam splitters with transmissivity $\eta$ modeling photon loss. }
\label{interferometer}
\end{figure}

In the case of two-photon states, this is enough to obtain arbitrary symmetric states, so this is sufficient to obtain the optimal states.
In contrast, for the four-photon states we would need two independent parameters.
As there is only one parameter, $\chi$, which can be varied, we cannot obtain the exactly optimal states of the form
\begin{equation}\label{eopt}
\ket\psi_{\rm ex} \propto \ket{0,4} + \chi'_1\ket{1,3} + \chi'_2\ket{2,2} + \chi'_1\ket{3,1} + \ket{4,0} .
\end{equation}
where $\chi'_1$ and $\chi'_2$ are independent real variables. However we will show that we can obtain results close to optimum.
In the following sections we show the effect of photon loss on states of the form \eqref{lr} in an adaptive phase measurement scheme shown in Fig.~\ref{interferometer}.

\section{Photon loss}\label{section-photonloss}
When using nonclassical states in optical interferometers, photon loss is the most important source of decoherence.
A standard way to model photon loss in these systems is to introduce fictitious beam splitters in the arms of the interferometer \cite{Lee}.
Here we consider the same amount of loss in both arms.
The transmissivity, $\eta$, of the fictitious beam splitters determines the efficiency of the system.

The loss-resistant states in Eq.~\eqref{lr} can be written in the form
\begin{equation}
\ket\psi_{\rm input} = \sum\limits_{k = 0}^N \psi_k\ket{N - k,k},
\end{equation}
where $\psi_k$ are the amplitudes of the photon number components, and depend on the single parameter $\chi$.
Considering the fictitious beam splitters after the phase shifts in the arms of the interferometer we define $\Psi_k=\psi_k e^{i(N - k)\varphi}e^{ik\theta}$.
The effect of the fictitious beam splitters is to change the creation operators of the upper and lower arms, $\hat b^\dag_1, \hat b^\dag_2$, to $\sqrt \eta  \hat b^\dag_1  +  i\sqrt {1 - \eta } \hat c^\dag_1 $  and $\sqrt \eta  \hat b^\dag_2  + i\sqrt {1 - \eta } \hat c^\dag_2 $, where $\hat c^\dag_1$ and $\hat c^\dag_2$ are the loss modes.
Thus, after loss the state $\ket{N-k,k}$  becomes
\begin{align}
&  \sum_{n = 0}^{N - k} \sum_{m = 0}^k i^{n + m}\sqrt {C_n^{N - k}C_m^k} \left(\sqrt \eta \right)^{N - n - m}\left( \sqrt {1 - \eta }\right)^{n + m}\nn
 &\qquad\times \ket{N - k - n,k - m,n,m},
\end{align}
where the third and fourth modes in the state are the loss modes and $C_m^k=\binom{k}{m}$ is the binomial coefficient.
If we denote the total number of photons lost as $L=n+m$, and trace over loss modes, the density operator can be written as
\begin{align}
\rho  &= \sum\limits_{L = 0}^N \sum_{m = 0}^L \sum_{r,s = 0}^{N - L} \Psi_{r + m}\Psi_{s + m}^* A_{N,L,r,m}A_{N,L,s,m}^* \nn
&\qquad\times\ket{N-L-r,r}\bra{N-L-s,s},
\end{align}
where 
\begin{equation}
A_{N,L,r,m} = \sqrt {\eta ^{N - L}(1-\eta)^L C_{N - L - r}^{N - r - m}C_r^{r + m}} .
\end{equation}
\begin{widetext}
After loss, a 50/50 beam splitter acts on the state. Calculating the output density operator, its diagonal elements give the output detection probabilities
\begin{align}\label{Plk}
P_{L,k}(\varphi ,\theta) &= \sum_{m = 0}^L \sum_{r,s = 0}^{N - L} \sum_{{r_2} = \max (0,k - N + L + r)}^{\min \left( {r,k} \right)} \sum_{{s_2} = \max (0,k - N + L + s)}^{\min \left( {s,k} \right)} \Psi_{r + m}\Psi_{s + m}^* A_{N,L,r,m}A_{N,L,s,m}^* \left( \frac 12 \right)^{N - L}(-1)^{r_2+s_2}\nn
& \quad \times \frac{(N-L-k)!k!}{\sqrt{(N-L-r)!r!(N-L-s)!s!}}
\binom{N - L - r}{k - {r_2}}
\binom r{r_2}
\binom{N - L - s}{k - {s_2}}
\binom s{s_2},
\end{align}
\end{widetext}
where $L$ is the total number of photons lost and $k$ is the number of photons detected in one of the output ports (which can be from $0$ to $N-L$).
For the two-photon input state, Eq.~\eqref{N2}, there are six different output probabilities:
\begin{itemize}
\item if no photons are lost ($L=0$), then we can detect zero, one and two photons in output mode 2 ($k=0, 1, 2$),
\item if one photon is lost ($L=1$), then we can detect zero or one photon in output mode 2 ($k=0, 1$), and
\item if both photons are lost ($L=2$), then the only possible detection result is vacuum ($k=0$).
\end{itemize}
Similarly for the four-photon state, Eq.~\eqref{N4}, there are fifteen output probabilities.
The larger number of output probabilities in comparison to the lossless case makes the calculations more computationally difficult.

Because the probabilities depend on the $\Psi_k$ coefficients, which in turn contain exponentials of the phase $\varphi$, the probabilities $P_{L,k}(\varphi ,\theta)$ can be written as a Fourier series 
\begin{equation}
\label{four}
P_{L,k}(\varphi ,\theta)= \frac 1{2\pi}\sum_j a_j e^{-ij\varphi}.
\end{equation}
When updating the phase probability by Bayes' theorem, the probability can be represented by a finite number of the Fourier coefficients.

In the following section we describe how our proposed states can be utilized in an adaptive measurement scheme to obtain phase variances less than the SQL.
\section{The measurement scheme}\label{section-adaptive}

The measurement scheme we propose is a sequence of states, produced as in Fig.~\ref{thescheme}, input to a lossy interferometer as in Fig.~\ref{interferometer}.
We combine the results of detection of each of these successive states to provide an overall measurement of the phase that is unambiguous, in a similar way as Refs.~\cite{HigginsN,DominicPRA9,NJPHiggins,Pezze}.
We use the terminology ``detection'' for the measurement of an individual state, to contrast with the overall measurement of the phase combining results of the individual detections.

Moreover, we use feedback to adjust the controlled phase $\theta$ based on the previous detection result and controlled phases.
In previous work this approach is usually found to give improved performance.
The globally optimal controlled phase is the one that minimizes the final phase variance, but finding such a controlled phase requires a minimization over an exponentially large number of variables.
Here we adopt the approach of finding the locally optimal phase, that minimizes the variance in the phase estimate after the next detection \cite{DominicBreslin,Xiang,DominicPRA09}.
There are proposals to the globally optimal phase in a more restricted sense that avoids needing an exponential number of variables \cite{Hentschel10,Hentschel11,Hayes}, but those are still more computationally intensive than finding the locally optimal phase.

The measure we use for the variance of the phase is the Holevo phase variance, $V_{\rm H} = \mu^{ - 2}-1$, \cite{Holevo} where $\mu$, called sharpness, is
\begin{equation}\label{sharpness}
\mu = \left| \left\langle e^{i\varphi } \right\rangle \right|.
\end{equation}
The Holevo phase variance coincides with the usual variance for distributions sharply peaked well away from the phase cut.

The Cram\'er-Rao inequality sets a lower bound for the phase variance of an unbiased measurement \cite{Braunstein}
\begin{equation}
V \ge \frac 1{F(\varphi ,\theta)}
\end{equation}
where 
\begin{equation}\label{Fisherinfo}
F\left( {\varphi ,\theta } \right) = {\sum\limits_{{u_k}} {\frac{1}{{P\left( {{u_k}|\varphi ,\theta } \right)}}\left( {\frac{{\partial P\left( {{u_k}|\varphi ,\theta } \right)}}{{\partial \varphi }}} \right)} ^2}
\end{equation}
is the Fisher information.
Note that this is for the usual variance (root-mean-square error), rather than the Holevo variance.
The Fisher information effectively represents the amount of information about $\varphi$ which is contained in the measurement result (though it is not quantified in bits as in the case of entropy).
The probability $P(u_k|\varphi ,\theta)$ is the probability of the measurement result $u_k$ given the system and controlled phases $\varphi$ and $\theta$.
It is the probability given in Eq.~\eqref{Plk}.

\begin{figure}[t!]
\centering
\includegraphics[scale=0.69]{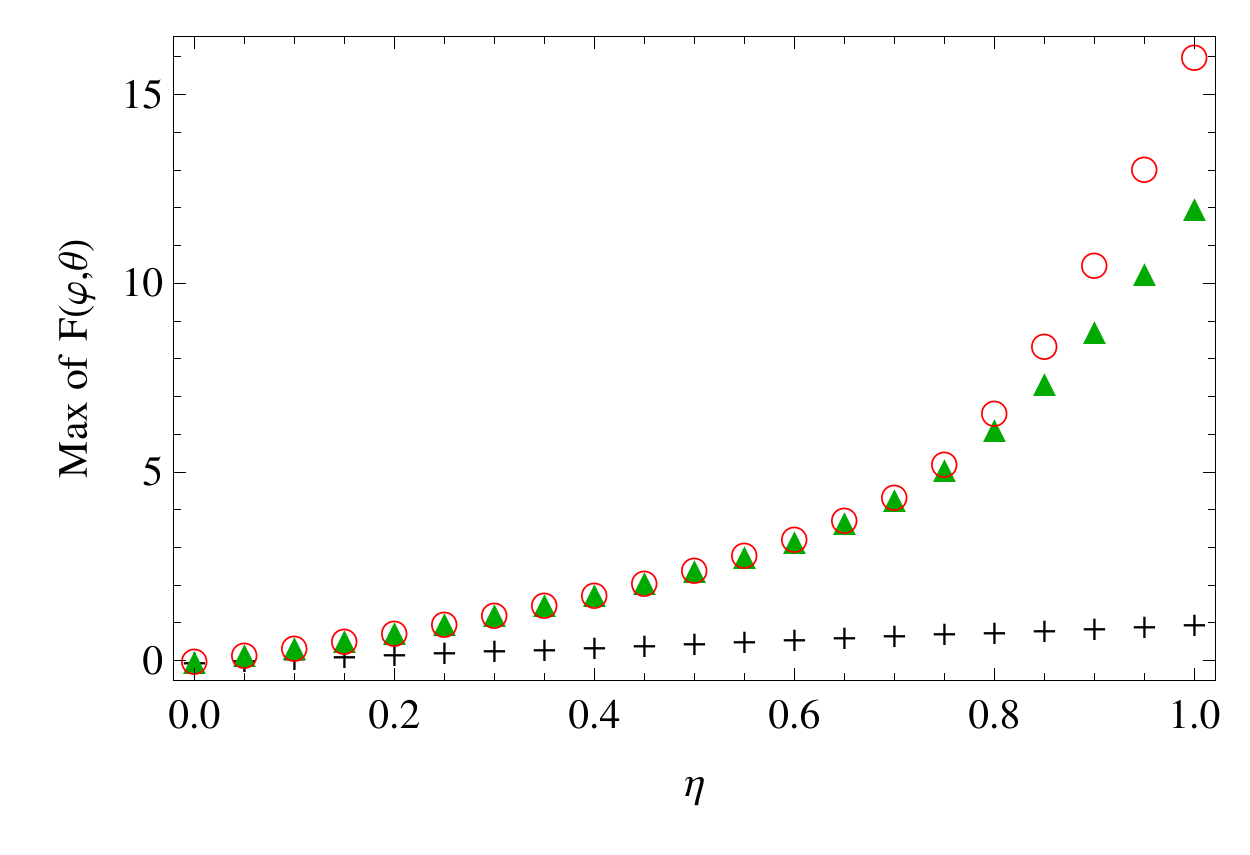}
\caption{Maximum of Fisher information $F(\varphi,\theta)$ versus efficiency $\eta$ for $\theta=0$. Circles: four-photon exact optimal state, Eq.~\eqref{eopt}. Triangles: four-photon loss-resistant optimal states, Eq.~\eqref{N4}. Pluses: single-photon state.}
\label{FI3}
\end{figure}

Optimal states for phase measurement with loss are typically evaluated via the Fisher information \cite{Dorner,Dobrzanski,Knott,Kacprowicz}.
Using the Fisher information to evaluate our proposed four-photon loss resistant states as given in Eq.~\eqref{N4}, they are nearly as good as the general optimal states.
As is shown in Fig.~\ref{FI3} the results are almost indistinguishable for $\eta<0.7$.

\begin{figure}[t!]
\centering
\includegraphics[scale=0.96]{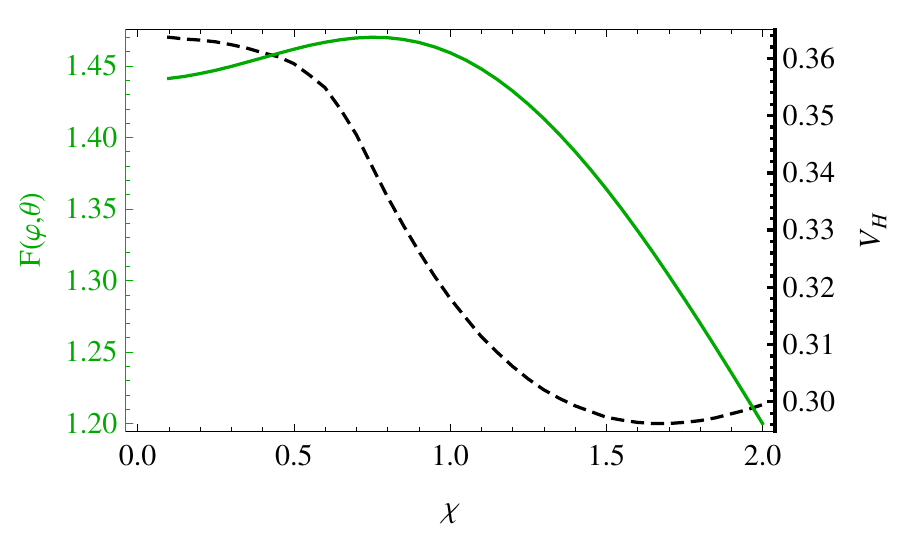}
\caption{Solid line (green): Fisher information for $\varphi  = \pi /4$, $\theta  = 0$ versus $\chi$ calculated using Eq.~\eqref{Fisherinfo} for the two-photon state given in Eq.~\eqref{N2}. Dashed line (black): phase variance versus $\chi$ for the sequence of seven single photons followed by one two-photon loss-resistant state in the adaptive measurement protocol.
The Fisher information gives a lower bound to the phase variance, rather than an exact phase variance, so the value of $\chi$ that maximizes the Fisher information need not be the value that minimizes the phase variance.}
\label{FI}
\end{figure}

However, the Fisher information only provides a lower bound on the variance.
To provide a better test of our states, we consider the phase variance produced by measurements.
The scheme we propose for phase estimation is as follows.
We use a sequence of $N_1$ single photon states followed by $N_2$ two- and $N_4$ four-photon loss-resistant states given in Eqs.~\eqref{N2} and \eqref{N4}.
The value of $\chi$ must be determined for each kind of state for each amount of photon loss in the system.
We have found the value of $\chi$ by numerical search over the states that minimize the phase variance.
For $\eta=0.6$ we found the optimal value of $\chi$ for two-photon states to be $1.7$ or $1.8$, while for four-photon states we found it to be $1.3$ (for a total number of photons up to $N=30$).
These values do not maximize the Fisher information given in Eq.~\eqref{Fisherinfo}; an example is shown in Fig.~\ref{FI}.
In this figure the Fisher information is shown for the two-photon loss-resistant state for the system phase $\varphi  = \pi /4$ and controlled phase $\theta=0$, and the maximum is for $\chi\approx 0.8$.
We also show the phase variance of a sequence of seven single photons followed by one two-photon loss-resistant state.
That is, there is one state dependent on the value of $\chi$, and the remaining states are to resolve the phase ambiguity.
In this case the minimum is for $\chi\approx 1.7$, which is a radically different value to that which maximizes the Fisher information.

We use Bayes' theorem to update the probability of the system phase given the measurement results $u_m$ and controlled phases $\theta_m$
\begin{equation}
P( \varphi |\vec u_m,\vec \theta_m ) \propto P( u_m|\varphi , \theta_m)P(\varphi |\vec u_{m - 1},\vec \theta_{m - 1} )
\end{equation}
where $\vec u_m=(u_1,u_2,\ldots,u_m)$ is the vector of successive measurement results and $\vec \theta_m=(\theta_1,\theta_2,\ldots,\theta_m)$ is the vector of the corresponding controlled phases (i.e.~$u_j$ is the measurement result with controlled phase $\theta_j$).
We also adopt the notation that $\vec u_0$ and $\vec\theta_0$ are the empty vectors.
The proportionality factor is just a normalization constant, which is trivial to calculate.
We assume that the phase is initially unknown, so the initial probability distribution is flat, and $P(\varphi |\vec u_0,\vec\theta _0)= 1/2\pi$.
The probability $P(u_m|\varphi ,\theta_m)$ is that given in Eq.~\eqref{Plk}.

We set the first controlled phase to zero.
This gives the same result as a random phase, because we average over the system phase and only the relative phase between the arms is significant.
The other controlled phases are obtained by maximizing the sharpness after the next detection.
In particular, the optimal $\theta_m$ is the one that maximizes
\begin{equation}
\mu(\theta_m) = \frac{1}{2\pi}\sum_{u_m} \left| \int_0^{2\pi} d\varphi \, e^{i\varphi}\prod_{k = 1}^m P(u_k|\varphi,\theta_k)\right|,
\end{equation}
where the summation is over all the possible results for the $m$'th measurement, and the product over $k$ corresponds to the updates of the probability distribution according to Bayes' rule.
Because the probabilities are represented as a Fourier series, as in Eq.~\eqref{four}, the integral over $\varphi$ simply yields the coefficient $a_{-1}$.
Therefore this sum may be obtained by summing the predicted values of $|a_{-1}|$ after the next detection.

For the case of measurement with a single photon without loss, the formula for the controlled phase from Ref.~\cite{DominicBreslin} may be used.
In the presence of loss the only extra detection result is where the photon is lost with a probability that is independent of the system and controlled phases.
This just adds an extra constant to the sharpness and does not change the phases that maximize it.
Therefore the formula from Ref.~\cite{DominicBreslin} may still be used.
That is, we take one of the following three phases
\begin{equation}
\theta_0 = \arg \left(ba^* - c^*a \right),\quad \theta_\pm = \arg \left( \sqrt{\frac{c_2 \pm \sqrt {c_2^2 + |c_1|^2} }{c_1}} \right),
\end{equation}
where 
\begin{align}
c_1 &= \left( a^*c \right)^2 - \left( ab^* \right)^2 + 4\left( |b|^2-|c|^2 \right)b^*c, \nn
c_2 &=  - 2i\mathop{\rm Im} \left( a^2 b^* c^* \right),
\end{align}
and  $a, b$ and $c$ are functions of Fourier coefficients for the probability distribution: $a = a_{-1}$, $b = \frac 12 a_{-2}$, $c = \frac 12 a_0$.
The optimal phase out of the above three possible phases is determined numerically.

The situation is more complicated with two- and four- photon loss-resistant states.
For the simpler states considered in Ref.~\cite{Xiang}, it was possible to use the above formula in the two-photon case.
However, here we have the complication that there is an additional $\ket{1,1}$ term in the state, and we also need to take account of the case where one photon is lost.
This means that the formula no longer applies.
For this reason we determined the optimal controlled phase numerically for the two- and four-photon cases.

We have calculated the exact phase variance by considering all the possible measurement results, of which there are $3^{N_1} \times 6^{N_2} \times 15^{N_4}$, (considering loss in all parts of the sequence).
We sum over the sharpness for each sequence of measurement results as
\begin{equation}
\mu = \frac{1}{2\pi}\sum_{\vec u_m} \left| \int_0^{2\pi} d\varphi \, e^{i\varphi}\prod_{k=1}^m P\left(u_k|\varphi,\theta_k\right) \right|.
\end{equation} 
In this expression we also average over the system phase in order to obtain the average performance of the scheme.
See Ref.~\cite{DominicPRA9} for a discussion of the theoretical basis for this approach.

To speed up the calculations it is useful to note that a measurement where $n$ of the single photons were not lost is the same as one where there were $n$ single photons without loss.
This is because, if a single photon is lost there is no phase information and the probability distribution and controlled phase need not be updated.
Using $\tilde\mu_n$ to denote the sharpness resulting from the sequence $n N_2 N_4$ with no loss on the single photons,
the actual sharpness of the sequence $N_1 N_2 N_4$, when there is loss in all parts, is given by
\begin{equation} 
\mu = \sum_{n=0}^{N_1} \binom{N_1}{n}\eta^n(1-\eta)^{N_1-n}\tilde\mu_n.
\end{equation}
In the above equation, $\binom{N_1}{n}\eta^n(1-\eta)^{N_1-n}$ is the probability of losing $n$ photons out of $N_1$ single photons.
If we performed the calculation in the obvious way, where there are three possible measurement results for each single photon, the number of measurement results to sum over for the single photons would be $3^{N_1}$.
By performing the calculation in this way, the number needed is $1+2+\ldots+2^{N_1}=2^{N_1+1}-1$, which is considerably less.

It is possible to consider arbitrary sequences of states, where the one- two- and four-photon states are used in any order.
To simplify the range of possible sequences to search over, we grouped together states of the same photon number.
A similar approach was used in Ref.~\cite{Xiang} for the lossless case, where numerical testing with small total numbers of photons found that this was optimal. Note that due to the adaptive nature of the measurement scheme the order of one, two- and four-photon states is important. 
We performed numerical searches over possible sequences ${N_1}{N_2}{N_4}$ (single photons followed by two- and four-photon loss-resistant states) to find the sequences which give the least variance for a range of total photon numbers.
The sequence configurations for some of the total numbers of photons are given in the table below.
\begin{center}
{\renewcommand{\arraystretch}{1.4}
\begin{tabular}{|c|c|c|c|c|c|}
\hline
$N$ & $N_1$ & $N_2$ & $\chi_2$& $N_4$   & $\chi_4$ \\
\hline
9 & 7 & 1&  1.7 & 0 &- \\
\hline
13& 7 & 1 &1.7 &1  &1.3  \\
\hline
30& 2 & 2 &1.8& 6 &1.3  \\
\hline
\end{tabular}}
\end{center}
where $\chi_2$ and $\chi_4$ are the values of $\chi$ for two- and four-photon states respectively.
\begin{figure}[t!]
\centering
\includegraphics[scale=0.7]{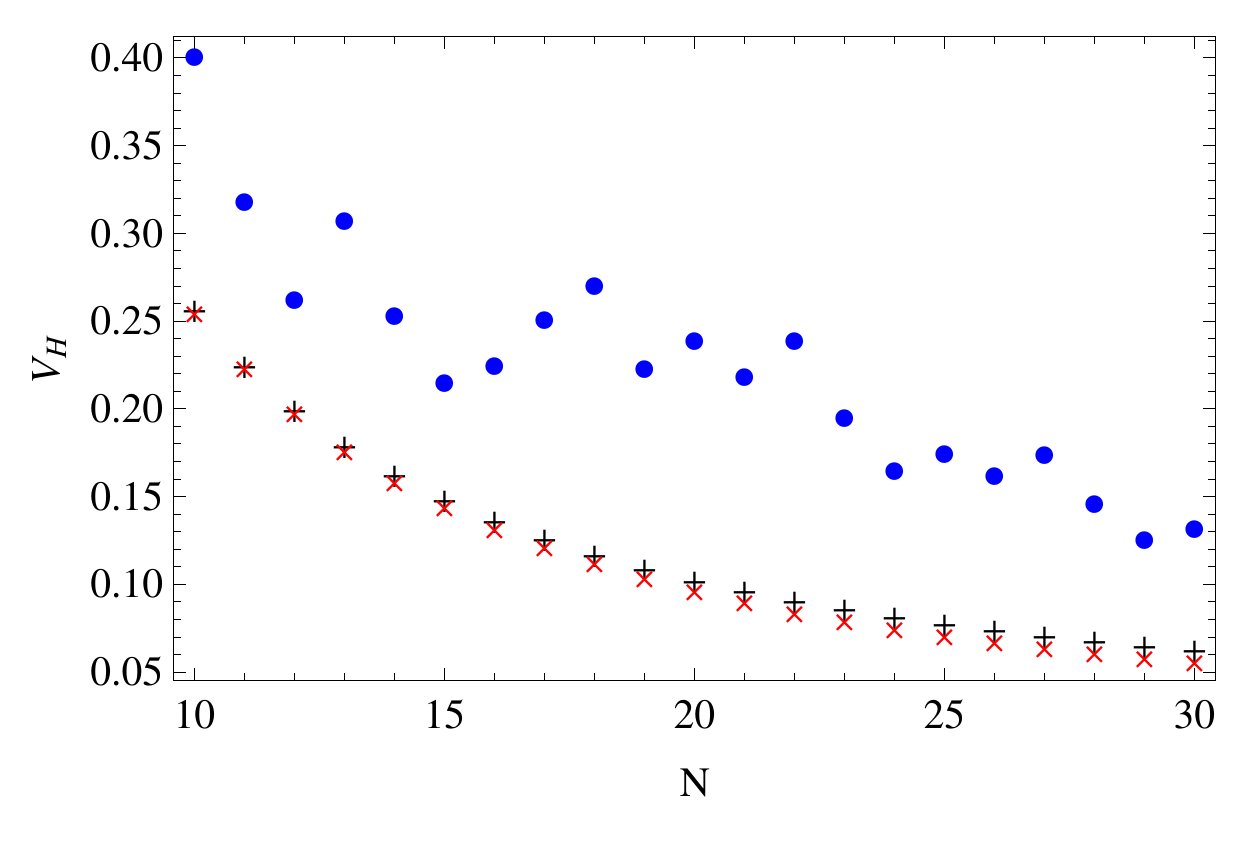}
\caption{The phase variance $V_{\rm H}$ versus total number of photons $N$ for $\eta=0.6$ for three different input states.
Plusses: only single photon states (SQL).
Crosses: optimal sequence of single photon states combined with two- and four-photon loss-resistant states.
Circles: the scheme proposed in \cite{Xiang}.}
\label{variance}
\end{figure}

The results of the numerical optimization for $\eta=0.6$ are plotted in Fig.~\ref{variance}.
In this figure the result for just single-photon states (with $\eta=0.6$) is shown for comparison, and can be regarded as equivalent to the SQL. 
Up to a total number of nine photons there is no advantage in using optimized multiphoton states, but after that the optimal sequence beats the SQL.
In this figure we have also shown the variance calculated using the states and sequences proposed in \cite{Xiang}.
The two- and four-photon states proposed in \cite{Xiang} are obtained by setting $\chi=0$ in Eqs.~\eqref{N2} and \eqref{N4}.
These states give variances which are considerably larger than those obtained from single-photon states.

This seems surprising, because using a sequence of states with different numbers of photons should be able to outperform a scheme limited to single-photon states.
However, that requires choosing the values of $N_1$, $N_2$, and $N_4$ appropriately, and the order that the states are used.
In this case we have considered the scheme of \cite{Xiang} using the values of $N_1$, $N_2$, and $N_4$ that were chosen to minimize the variance \emph{without} loss.
It is clear that the optimal values of these quantities must be dependent on the loss;
that is, the improvement over the scheme of \cite{Xiang} is primarily due to choosing the state sequences in such a way as to optimize the measurements for loss.
 
\section{Conclusion}\label{ch6}
Here we proposed an approach to generate multi-photon entangled states which are optimal for phase measurement in the presence of photon loss.
In order to provide a technique that is experimentally feasible, we have considered methods of processing SPDC sources in order to provide improved loss tolerance.
For two-photon states the method produced optimal states, but the technique is not able to exactly produce the optimal four-photon states.
However, as shown in Fig.~\ref{FI3}, the maximum of Fisher information for our proposed four-photon loss-resistant states is almost the same as the exact optimal states up to $\eta=0.7$.

We have proposed techniques of combining these loss-resistant states in order to provide an unambiguous measurement of the phase.
Surprisingly, we find that the parameters that minimize the phase variance are not the same as those that maximize the Fisher information.
This is likely because the Fisher information is more relevant in the asymptotic case of large numbers of copies of the state, whereas we are considering a small number of copies.

By using the sequence of loss-resistant states, we are able to beat an SQL defined by the corresponding scheme with independent single photons.
In comparison, if the measurement scheme of Ref.~\cite{Xiang} is used, the phase variance is much greater.
In order to obtain the best performance, the state sequence should be chosen based on the loss.
Optimizing the parameters for the loss-resistant states will provide an additional improvement.

\section{Acknowledgements}
We thank Geoff Pryde for helpful discussions. DWB is funded by an ARC Future Fellowship (FT100100761).

\end{document}